# Single mode waveguide platform for spontaneous and surface-enhanced on-chip Raman spectroscopy

Ashim Dhakal[1,2*], Frédéric Peyskens[1,2*], Stéphane Clemmen[1,2], Ali Raza[1,2], Pieter Wuytens[1,2,3], Haolan Zhao[1,2], Nicolas Le Thomas[1,2] and Roel Baets[1,2]

[1]*Photonics Research Group, INTEC Department, Ghent University–imec, Belgium*
[2]*Center for Nano- and Biophotonics, Ghent University, Belgium*
[3]*Department of Molecular Biotechnology, Ghent University, Belgium*
* Contributed equally to the work

We review an on-chip approach for spontaneous Raman spectroscopy and Surface Enhanced Raman Spectroscopy (SERS) based on evanescent excitation of the analyte as well as evanescent collection of the Raman signal using Complementary Metal Oxide Semiconductor (CMOS) compatible single mode waveguides. The signal is either directly collected from the analyte molecules or via plasmonic nanoantennas integrated on top of the waveguides. Flexibility in the design of the geometry of the waveguide, and/or the geometry of the antennas, enables optimization of the collection efficiency. Furthermore the sensor can be integrated with additional functionality (sources, detectors, spectrometers) on the same chip. In this paper, the basic theoretical concepts are introduced to identify the key design parameters and some proof-of-concept experimental results are reviewed.

## I. Introduction

Raman scattering is a process where incident photons are scattered inelastically by vibrations characteristic to the scattering molecules [1-3]. The frequency of the scattered photon is shifted by an amount corresponding to vibrational energies of the molecules and the intensity of the scattered light is proportional to the number of scattering molecules. Hence, the Raman spectrum constitutes a specific pattern that allows one to identify and quantify the number of molecules. Thanks to its non-invasive and label-free nature, Raman spectroscopy has been established as a powerful spectroscopic technique for basic research and a myriad of applications ranging from biology, chemistry to material sciences [2,3]. Scientists have applied Raman spectroscopy techniques for a wide variety of remarkable applications such as the study of viruses [4], the classification of tumour cells [5] and a single molecule detection [6].

Despite the enormous progress made over the past decades in demonstrating and identifying several applications, Raman spectroscopy has not yet found a widespread use as a point-of-need tool outside of the dedicated labs. The major drawback of Raman spectroscopy is the extremely small cross-section of the spontaneous Raman scattering process. It is in the order of $10^{-30}$ cm$^2$, which is fifteen orders of magnitude smaller compared to fluorescence [1]. The extremely weak nature of the Raman scattering process necessitates advanced techniques to enhance and detect the signal. Coherent Raman scattering [7], stimulated Raman scattering, and surface enhanced

Raman scattering [8, 9] (SERS) are only a few examples of techniques devised to enhance the Raman signal. However, in conventional setups these techniques still require a confocal microscope combined with advanced laser sources, detectors and a large monochromator, hence, limiting the use of the Raman spectroscopic techniques to lab environments.

The development of optoelectronic and photonic technologies such as diode lasers and CCD detectors has enormously advanced the use of Raman spectroscopy for numerous applications [3]. Recent advances in integrated photonics technology have opened up unique ways towards cheap and compact point-of-need Raman analysis tools by integrating the essential components of the spectroscopic system on a chip [10, 11]. As for many other optical systems, a Raman spectroscopy system may be miniaturized and integrated on a chip. Indeed, all the essential components such as lasers [12], detectors, planar concave gratings [13], arrayed waveguide gratings [14, 11], Mach-Zehnder interferometer filters [15, 16], ring cavity filters [17] and Bragg gratings [18] have been demonstrated using chip-scale nano-photonics. However, effective transduction methods to generate Raman signals in an on-chip integrated Raman spectroscopic system have remained largely unexplored. A large enhancement and a small étendue (the product of the area and the solid angle of a light beam) are two key requirements of a transduction method for an integrated Raman sensor. A large enhancement is needed to eliminate the necessity of cooled detectors and expensive filters which are typically used in Raman spectroscopy. A small étendue is essential to keep the size of the integrated spectrometer as small as possible for a given resolution. These requirements have neither been identified nor addressed in an integrated setup.

Nevertheless, Kanger et al, have demonstrated the use of multilayer planar waveguides for the evanescent excitation of adsorbed thin layers for Raman spectroscopy [19]. Although this technique is not readily integrable with an optimal on-chip spectrometer, and still requires a conventional microscope to collect the signal, this work took the idea of confining light and extending the interaction volume to boost the Raman signal using waveguides proposed by Rabolt et. al. [20]. We refer elsewhere [21] for a review of spectroscopic techniques based on planar waveguides. Photonic crystal fibres filled with gaseous analyte [22] and hollow core fibres filled with liquid analyte [23] are key examples of the techniques that recognize wave-guiding as a mechanism to enhance Raman signal, but fail to be directly integrable on a chip.

We have recently shown that single mode waveguides can be used to evanescently excite and collect Raman signals and have intrinsic assets to implement Raman spectroscopy in a lab-on-a-chip framework [24]. Waveguides combine a large detection volume with the field enhancement near a high-index-contrast waveguide to enhance the collected Raman signal [25]. Furthermore, the étendue of the collected signal in a single mode waveguide is by definition the smallest, thus allowing integration with the smallest possible integrated spectrometers for a given spectral resolution. A high performance, low cost, and compact Raman sensor can thus be integrated on a chip. Further, SERS can be performed by integrating nanoplasmonic antennas [26, 27] on the single mode waveguides when a considerably higher signal enhancement is desirable for the detection of small volumes of analyte. Here we review the use of single mode dielectric waveguides, which are possibly functionalized with nanoplasmonic antennas, for on-chip spontaneous and enhanced Raman spectroscopy.

This article is structured as follows. First, we discuss the collection efficiency of an ideal conventional Raman confocal microscope and identify its limitations. Subsequently, we introduce the silicon nitride waveguide platform which is used for our on-chip Raman work. Beginning with a discussion of planar waveguide enhancement in the fourth section, we discuss and calculate the overall evanescent collection efficiency for single mode channel waveguides. In the fifth section, we discuss the prospects of on-chip integration of plasmonic nanoantennas for SERS and identify the key figure of merit. The sixth section is dedicated to the experimental verification of the concepts discussed in the earlier sections. A brief comparative analysis of spontaneous and enhanced on-chip Raman spectroscopy is included in the seventh section, before concluding in the eighth section.

## II. Collection efficiency of diffraction-limited optics

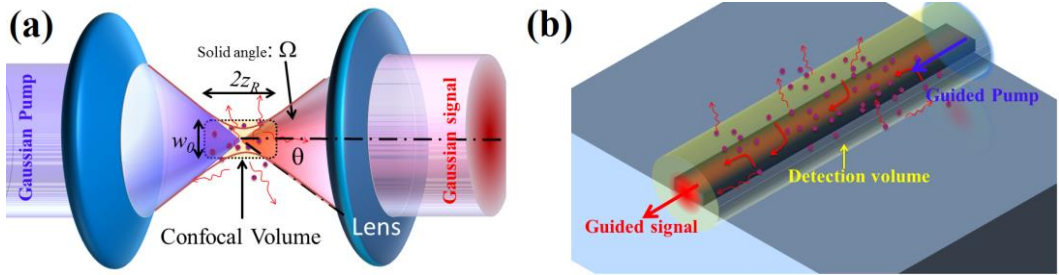

Fig 1. (**a**) Schematic of the beam geometry for a confocal microscope. Only signal from confocal volume contributes significantly to the signal. (**b**) Schematic of the waveguide based evanescent Raman sensors. A long interaction length leads to a very large detection volume.

In case of confocal microscopes employing diffraction limited excitation and collection, the collected signal essentially originates from particles located in the confocal volume, i.e. within the depth-of-focus $2z_R$, where $z_R$ is the Rayleigh range of the beam. Further, only the emission inside a solid angle $\Omega$ corresponding to a half-acceptance angle $\theta$ (Fig 1) of the collection optics is collected. We assume that collection optics has the same acceptance width (slit width of the spectrometer) as the $1/e^2$ width of the exciting beam, such that the half-acceptance angle is the same as that of the divergence of the input beam. In terms of differential scattering cross-section $\sigma$ and the density of the particles $\rho$, the overall efficiency of excitation and collection can then be calculated as [**28**]

$$\frac{P_{DL}}{P_{pump}} = 2z_R \Omega \sigma \rho \qquad (1)$$

Assuming a small beam divergence, $\theta = w_0/z_R = \lambda_0/(n\pi w_0)$, where $w_0$ is the beam waist at the focus, the solid angle can be expressed in terms of the parameters of the Gaussian beam:

$$\Omega = 4\pi \sin^2\left(\frac{\theta}{2}\right) \approx \pi \left(\frac{w_0}{z_R}\right)^2 = \frac{\lambda_0}{nz_R} \qquad (2)$$

Hence,

$$\frac{P_{DL}}{P_{pump}} = 2\left(\frac{\lambda_0}{n}\right)\sigma\rho \qquad (3)$$

Note that the SI unit for solid angle steradian (sr) is implicit in the pre-factor of Eq. (3). The approximation carried out in Eq. (2) overestimates the solid angle, as in general, $w_0/z_R > \sin(\theta)$. Hence, Eq. (3) sets the upper-limit for the confocal microscopic systems. Microscopes with higher transmission losses and having larger divergence than diffraction-limited systems behave worse. Eq. (3) is independent of any system-specific parameters and reveals the interplay between the beam divergence and the depth of focus, which restricts the detection volume and field intensity, hence, the collected power [**29**]. In the coming sections, we describe the use of dielectric waveguides to overcome the limit imposed by diffraction by means of waveguiding and electromagnetic confinement. The next section briefly introduces dielectric waveguides and the relevant technology platform for the fabrication of such waveguides.

### III. Silicon nitride integrated waveguides for Raman spectroscopy

Dielectric waveguides constitute a high refractive index ($n_{core}$) material called *core* surrounded by, one or more, lower refractive index materials called *cladding* (with lowest refractive index $n_{clad}$) so that the light is guided in the core by total internal reflection [**30**]. Optical fibres are a well-known example of dielectric waveguides with a very low *index contrast* $\Delta_n$, defined as $\Delta_n = 1 - n_{clad}/n_{core}$. In the realm of integrated photonics, however, high index silicon (Si) core (refractive index $n_{si}=3.45$) material is defined on top of a silica ($SiO_2$) undercladding (refractive index $n_{ox} = 1.45$), by selectively etching the unwanted Si region [**31-32**]. This platform for defining Si waveguides on $SiO_2$ is popularly called the silicon-on-insulator (SOI) technology platform and it is the epitome of high index contrast waveguide integration platforms in a well-developed CMOS compatible fabrication technology. High index contrast waveguides confine the light in a very small area, which leads to many advantages. One important advantage of the high index contrast waveguide is the enhancement of the field intensity in the vicinity of the waveguides, thus increasing the light-matter interaction which we shall explore in section IV. Another critical advantage is the significant reduction of étendue of pump and signal which allows small spectrometers with an optimal spectral resolution and facilitates compact integration.

A standard SOI integration platform, with a pump wavelength $\lambda_0 = 1.55$ µm, would be a possible choice for on-chip Raman spectroscopy, owing to its maturity and extensive use in the long-haul telecom applications. However, the Raman scattering cross-section roughly scales as $\lambda^{-4}$ — inverse to the fourth power of the pump wavelength [**1**]. Thus, a shorter wavelength is generally preferred as long as the fluorescence and absorption, characteristically associated with the use of shorter wavelengths, remains manageable. A near-infrared wavelength of 785 nm is a popular choice in Raman spectroscopy for biological applications because of the low water absorption, low fluorescence originating from biological molecules, but still fairly short wavelength for achieving a reasonably high scattering cross-section. Availability of high quality and low-cost sources and detectors in the 700-1000 nm wavelength region is also an important drive for the choice of 785 nm wavelength. Unfortunately, the popular SOI platform is not suitable for 785 nm wavelength as silicon absorbs heavily for wavelengths < 1 µm, so an alternative integration platform is required.

Silicon nitride ($Si_3N_4$) is another common material used in CMOS fabrication technology, the backbone of modern electronics. It has a small absorption in the 500 nm – 2500 nm wavelength region, a moderately high refractive index $n_{sin}$ =1.8-2.2, exhibits low fluorescence and is a very robust material capable of handling large optical powers [33, 34]. Thus $Si_3N_4$ waveguides [35, 36] defined on top of a silica undercladding constitutes a good trade-off for Raman spectroscopy at visible and near-infrared wavelengths, satisfying several technological constraints.

$Si_3N_4$ waveguide circuits discussed in this article (except for the hybrid nanoplasmonic waveguides) are fabricated on 200 mm diameter silicon wafer with 200 mm diameter and thickness of 700 μm. A stack of 220 nm thick $Si_3N_4$ on top of 2.4 μm thick $SiO_2$ is deposited on top of the Si wafer. The $Si_3N_4$ as well as the $SiO_2$ are deposited by plasma enhanced chemical vapor deposition [35]. The waveguide structures are defined with 193 nm optical lithography and subsequently etched by fluorine based inductive coupled plasma-reactive ion-etch process to get the final structure. Fig 1(b) shows the structure of a typical $Si_3N_4$ waveguide.

## IV.    Enhancement and collection efficiency with waveguides

In this section we introduce waveguide enhancement with the help of planar waveguides and show the importance of index contrast for the waveguide enhancement. Next we calculate the overall evanescent collection efficiency for single mode channel waveguides, which are generally used in the context of integrated photonics.

Unless otherwise noted, the calculations are done for a wavelength $\lambda_0$ of 785 nm in case of $Si_3N_4$ and $\lambda_0$ = 1550 nm in case of SOI waveguides. The refractive indices of Si, $Si_3N_4$, $SiO_2$ and isopropyl alcohol IPA (analyte used as upper cladding) are respectively taken to be $n_{si}$=3.45, $n_{sin}$ = 1.89, $n_{ox}$ = 1.45 and $n_{ipa}$ = 1.37. The waveguide height $h$ is 220 nm.

**Enhancement of emission near dielectric waveguides**
To understand the enhancement effects inherent to the waveguides and the role of refractive index contrast, we investigate a common planar waveguide with a silica undercladding, an air uppercladding ($n_1$ =1), and variable core index $n_c$ and core thickness $d$. An emitting molecule nearby a waveguide can be modelled as a dipole oscillating in a certain direction at a given frequency [36, 37]. The total field density at the location of the molecule, which determines the emission, can be calculated by expanding the dipole field into plane wave components, and applying the electromagnetic boundary conditions [37, 38]. We investigate two common CMOS compatible core materials, Si ($n_c$ = 3.45) and $Si_3N_4$ ($n_c$ = 1.89), for the two orthogonal dipole orientations with respect to the slab waveguide surface. Fig. 1(a) shows the power $P$ coupled to the fundamental TE and TM modes of a slab waveguide from a dipole located at the surface of the waveguide, normalized to the total free-space emission $P_0$ of the dipole. Depending on the orientation of the dipole and polarization of the mode, a strong coupling of the dipole emission (even exceeding $P_0$) can be observed. The coupled power scales with the overlap of the dipole field and the modal field at the dipole location and is strongly dependent on the core thickness and core index. The same conclusions hold for the case of excitation from the waveguide modes [**25**, **39**]. Hence, in the absence of non-radiative transitions, the overall efficiency of excitation and collection by the waveguide is a quadratic function of the field intensity. These results from slab waveguides underscore the role of waveguide geometry and index contrast for efficient excitation and collection [**39**]. Interestingly, for both material systems at their respective wavelengths ($Si_3N_4$: 785 nm and Si: 1550 nm), the optimal slab thicknesses are near 120 nm and 240 nm, respectively for TE and TM excitation.

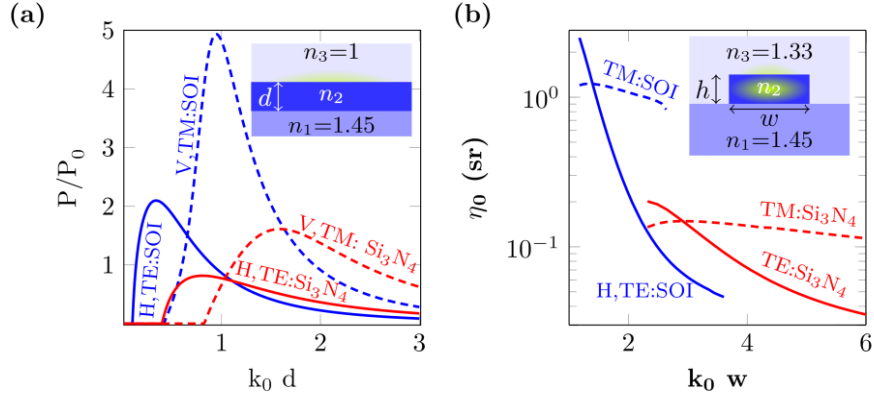

Fig. 2. (**a**) Normalized power $P/P_0$ coupled to the fundamental TE (solid) and TM (dashed) modes of a slab waveguide as a function of normalized waveguide width $k_0 d$, for Si (blue) and $Si_3N_4$ (red) cores. The powers are calculated for dipoles on the core surface and oriented vertical (V) and horizontal (H) to it. **Inset:** a generic slab waveguide (adapted from [**39**]) (**b**) Theoretical $\eta_0$ curves for strip waveguides as a function of the normalized waveguide width $w$ for Si (blue) and $Si_3N_4$ (red) waveguide systems for fixed $h$=220 nm. Solid lines: TE polarized excitation and collection. Dashed lines: TM polarized excitation and collection. Only the curve segments corresponding to a well-defined single mode operation are shown. **Inset:** a general strip waveguide (adapted from [**25**]).

**Conversion efficiency of channel waveguides**
The excitation efficiency of a particle located at a position $\mathbf{r_0}$ by a channel waveguide mode and the subsequent collection by the same mode can be described by a parameter called *integrated luminosity* $\Lambda_{wg}$ of the waveguide. It is defined as the ratio of collected power ($P_{wg}$) to pump power ($P_{pump}$) for a unit scattering cross-section ($\sigma$):

$$\Lambda_{wg}(\vec{r_0}) = \frac{P_{wg}(\vec{r_0})}{\sigma P_{pump}} \quad (4)$$

For a random ensemble of incoherently scattering particles with a uniform volume density $\rho$, distributed in a volume $V$, the total scattering efficiency can then be calculated as,

$$\frac{P_{Tot}}{P_{pump}} \equiv \sigma\rho \iiint_V \Lambda_{wg}(\vec{r_0}) d\vec{r} \quad (5)$$

In case of arbitrarily shaped channel waveguides, having a translational symmetry in the longitudinal direction transverse to the surface $S$, we define a quantity called *specific conversion efficiency* $\eta_0$ as:

$$\eta_0 \equiv \frac{d}{dz}\iiint_V \Lambda_{wg}(\vec{r_0}) d\vec{r} = \iint_S \Lambda_{wg}(\vec{r_0}) d\vec{r} \quad (6)$$

such that the power $P_{Tot}$ collected by a lossless waveguide section of length $l$ is given by,

$$\frac{P_{Tot}}{P_{pump}} = \eta_0 \rho \sigma l \tag{7}$$

Neglecting the Stokes shift, $\Lambda_{wg}$ can be written as [**25**],

$$\Lambda_{wg}(\vec{r}_0) = \frac{1}{n(\omega)} \left( \frac{n_g(\omega)\lambda_0 |\vec{e}_m(\vec{r}_0, \omega_p)|^2}{\iint_\infty \varepsilon(\vec{r}, \omega_p) |\vec{e}_m(\vec{r}, \omega_p)|^2 d\vec{r}} \right)^2 \tag{8}$$

where $n_g$ is the group index of the mode. Thus, the conversion efficiency is given by

$$\eta_0 = \frac{1}{n(\omega)} \iint_S \left( \frac{n_g(\omega)\lambda_0 |\vec{e}_m(\vec{r}_0, \omega_p)|^2}{\iint_\infty \varepsilon(\vec{r}, \omega_p) |\vec{e}_m(\vec{r}, \omega_p)|^2 d\vec{r}} \right)^2 \tag{9}$$

The conversion efficiency $\eta_0$ evidently depends on the dielectric function that defines the waveguide field distribution and can be calculated using standard mode solvers. The conversion efficiency $\eta_0$ has the unit of solid angle (sr), reminiscent of the collection efficiency of a diffraction-limited optical microscope described by Eq. (1).

In Fig. 2(b) we show the theoretical conversion efficiency $\eta_0$ calculated for $Si_3N_4$ and SOI strip waveguides for normalized waveguide widths $k_0w$ and fixed height $h$ =220 nm corresponding to single mode operation of the waveguides. As the waveguide width is increased, $\eta_0$ increases to a maximum corresponding to an optimal mode confinement in the core and the cladding. Once this condition is reached, further increase in the width leads to a decrease in $\eta_0$ owing to more confinement of the modal field in the core. This reduces the interacting field in the uppercladding region of the analyte. The discontinuity of the electric field at the core - top cladding interface of the TM modes leads to a higher evanescent field at the position of the analyte. Thus, higher index contrast waveguides with TM polarization generally perform better than TE polarized modes. The optimal widths for $Si_3N_4$ waveguides at 785 nm are near 260 nm and 300 nm for TE and TM excitation and collection respectively, while for SOI waveguides it is near 150 and 240 nm respectively.

**Slotted channel waveguides**
As depicted in Fig 3(a), slotted waveguides have a narrow slot of width *s* in between the rails of the core material of rail width *r*. Due to the boundary conditions of the electromagnetic field, the existence of a narrow slot leads to a huge enhancement of the field in the slotted region, and hence also to the value of $\eta_0$. For a fixed rail width *r* = 275 nm and thickness *h* =220 nm, Fig 3(a) shows $\eta_0$ calculated for $Si_3N_4$ and Si slot waveguides for variable slot widths using Eq. (9). A significant increase in $\eta_0$ is observed as *s* is reduced for TE modes owing to the enhancement of the TE-field in the slot region as *s* is decreased. In case of SOI waveguides, where the index contrast is higher and much higher field enhancement can be expected, the value of $\eta_0$ reaches up to 40 sr for *s* = 20 nm.

Fig 3(b) shows $\eta_0$ calculated for $Si_3N_4$ and SOI for a fixed slot width *s* = 150 nm and for various total widths *w*, inclusive of the slot width *s* (note: *r* = (*w-s*)/2). As observed in the case of strip waveguides, $\eta_0$ decreases when the waveguide width is increased as a result of the confinement of the mode in the core and consequent reduction of the interaction volume in the analyte region. Due to the continuity of the electric field, the field enhancement in the slot is much less in case of TM polarization leading to relatively insignificant enhancement in case of TM polarization.

**Equivalent length and enhancement**
Comparing Eq. (7) and Eq. (3), we can define an equivalent length $l_{eq}$, which corresponds to the length of the waveguide that gives the same signal as that of diffraction limited confocal microscopes.

$$l_{eq} = \frac{2}{\eta_0}\left(\frac{\lambda_0}{n}\right) \quad (10)$$

Depending on the index contrast, polarization and design of the waveguides, $\eta_0$ may typically vary from 0.1 sr to 50 sr. Thus a waveguide length $< 20\ \lambda_0$ wavelengths is enough to provide an efficiency equivalent to that of a diffraction-limited confocal microscope. A single mode waveguide loss of 1dB/cm is typical for the considered integrated photonics platform, allowing more than a centimeter of waveguide without significant loss of the pump or collected signal. This leads to 3-5 orders of magnitude enhancement of the signal compared to diffraction-limited systems.

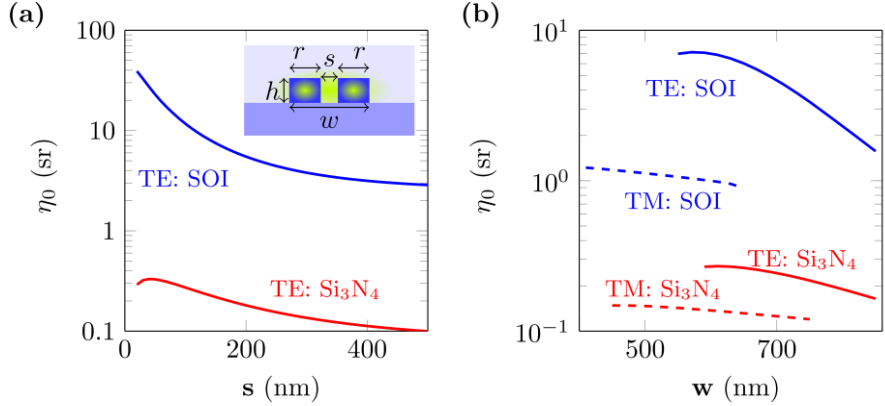

Fig. 3. Theoretical conversion efficiency $\eta_0$ for slotted waveguides for SOI (blue) and $Si_3N_4$ (red) waveguide systems. **a**) Fixed $r = 350$ nm, $h = 220$ nm, as a function of slot width $s$ for fundamental TE mode. **b**) Fixed $s = 150$ nm, $h = 220$ nm, as a function of waveguide width $w = s+2r$. Solid lines: TE polarized excitation and collection. Dashed lines: TM polarized excitation and collection. (adapted from [**25**])

**Lossy waveguides**
So far we have neglected the waveguide losses. Waveguide losses can be incorporated in our model by defining a function $\zeta(l)$ such that [**24**]:

$$\zeta(l) = \frac{P_{col}}{P_{tx}} = \frac{\eta_0 \rho \sigma}{2}\left(\frac{e^{\Delta\alpha l}-1}{\Delta\alpha}\right) \quad (11)$$

where, $P_{tx}$ is the transmitted pump power at the detector position and $\Delta\alpha$ is the difference in waveguide losses at the pump and Stokes wavelength.

## V. Hybrid nanophotonic-plasmonic waveguides

In the previous section we showed that nanophotonic waveguide based evanescent Raman spectroscopy provides a promising improvement in the sensitivity to standard confocal microscopy for bulk sensing. We showed that the most significant component of the enhancement is due to an increase in the interaction volume, and only a small component comes from the enhancement of the field itself. As a result, it becomes difficult to detect a small number of analyte molecules or nanoparticles in solution. Specifically designed metallic nanostructures allow a considerable enhancement of the electric field near the metal surface [**40**]. Such structures can hence locally boost the signal in order to allow detection of molecules with an extremely weak Raman cross-section or ultralow concentrations (i.e. SERS). SERS signals are however mainly excited and collected by bulky and expensive microscopy systems. The development of nanophotonic circuits functionalized with dedicated nanoplasmonic antennas would hence present a significant step towards the realization of dense SERS probes, allowing multiplexed detection of extremely weak Raman signals. Up till now, these integration efforts have been mainly limited to standard dielectric waveguides which are used to probe SERS signals from external, non-integrated, metallic nanoparticles [**40-44**]. Such an approach however loses quantitative prediction power due to the large uncertainty on the Raman enhancement and coupling between the waveguide mode and the metallic nanoparticles. In order to overcome this issue, we developed a fully integrated single mode SERS probe enabling on-chip excitation and emission enhancement in the 700-1000 nm region and developed an analytical model to outline the relevant design parameters and figure of merit for this new platform [**26, 27**].

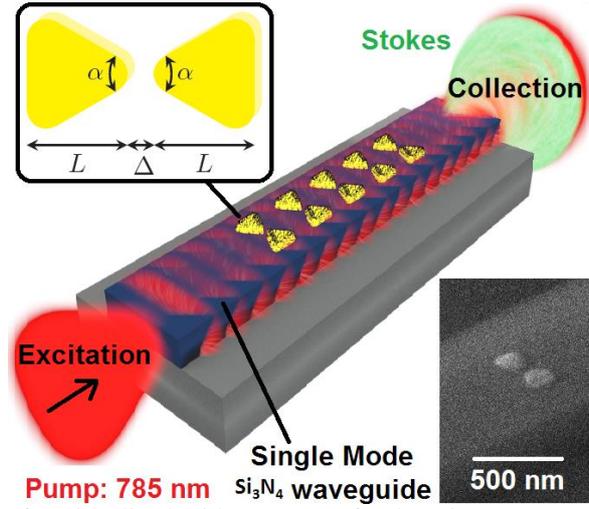

Fig. 4: $Si_3N_4$ waveguide functionalized with an array of $N$ bowtie antennas. The top inset depicts the geometrical parameters of a bowtie antenna (gap $\Delta$, length $L$ and apex angle $\alpha$). The bottom inset shows a SEM image of an integrated bowtie antenna (adapted from [**27**]).

A schematic of this on-chip SERS platform is shown in Fig. 4. The fundamental TE-mode of a $Si_3N_4$ rib waveguide excites a periodic array of $N$ gold bowtie antennas fabricated with a period $\Lambda$ [26]. A forward propagating Stokes power $P_s$ coupled to the waveguide, resulting from an analyte in the vicinity of an antenna, can be described in a manner similar to Eq. (7), with a quantity called the single antenna conversion efficiency $\eta_A$ defined as:

$$\frac{P_S}{P_{pump}} = \eta_{SERS} \rho \sigma = \eta_A \quad (12)$$

The single antenna conversion efficiency $\eta_A$ is an antenna dependent factor incorporating the integrated field enhancement profile near the metal surface as well as the analyte molecular density and Raman cross-section. This efficiency is related to the $\Lambda_{wg}$ defined in Eq. (4). In case of SERS, we define the conversion efficiency for the antenna-analyte system because the antenna enhancement is strongly affected by the molecular properties, such as its density and refractive index.

In addition to the intrinsic waveguide losses $\alpha_{wg}$, the pump and Stokes light will also be attenuated by the antenna extinction $e_p$ and $e_s$ respectively. It is shown [**26**] that the total forward propagating Stokes power $P_{tot}$ generated by an array of $N$ coated antennas is proportional to the following figure of merit *FOM*:

$$FOM(N) = \eta_A e_s^{1-N} \left( \frac{1 - \left(\frac{e_s}{e_p}\right)^N}{1 - \left(\frac{e_s}{e_p}\right)} \right) \tag{13}$$

The quantity *FOM* (*N*) contains all design parameters to assess the SERS signal strength for a given waveguide geometry and is therefore the proper figure of merit in comparing different antenna geometries.

In Figure 5 we compare $\eta_A$ and *FOM* (*N*) for four different bowtie antennas with fixed $\alpha = 60°$, $\Delta = 10$ nm, but varying length $L$ (70, 90, 110, 130 nm) and coated with a 1 nm 4-nitrothiophenol (NTP) layer.

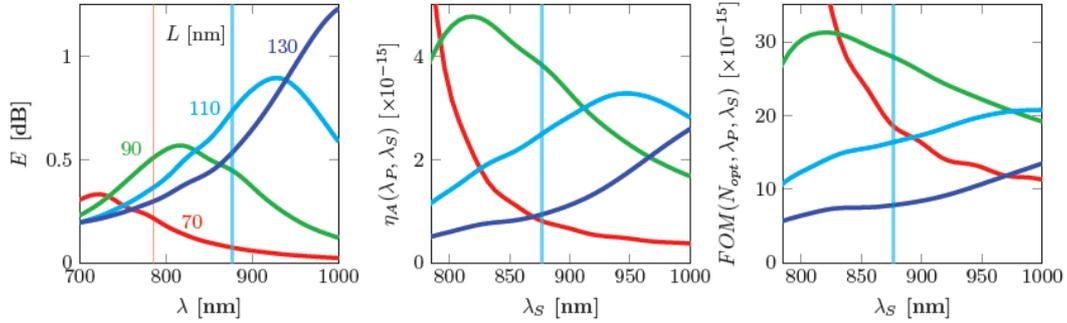

Fig. 5: Comparison of the antenna extinction $E$ (in dB), $\eta_A$ and *FOM* (*N*) for four different bowtie antennas with fixed $\alpha = 60°$ and $\Delta = 10$ nm but varying length $L$ (70, 90, 110, 130 nm). The red and cyan line mark the respective pump and Stokes wavelength (adapted from [**26, 27**]).

For a given pump and Stokes wavelength, *FOM* (*N*) is maximal for an optimum antenna number:

$$N_{opt} = \frac{\log\left(\frac{\log(e_s)}{\log(e_p)}\right)}{\log\left(\frac{e_s}{e_p}\right)} \tag{14}$$

For the selected Stokes wavelength (877 nm) from an NTP monolayer, a bowtie with length $L = 90$ nm generates the highest SERS signal as the *FOM* ($N_{opt}$) is maximal for this configuration. For this optimal bowtie geometry, a single antenna will produce about 4 fW of Stokes power for each 1 W of guided pump power as can be derived from the middle part of Fig. 5. Organizing these bowties in an array with $N_{opt}$ antennas will generate about 30 fW of Stokes power at the output for 1 W guided power. Similar calculations can be repeated for other geometries as well. In the experimental section we will show that the absolute Raman power predicted by this analytical model fits very well with the experimentally obtained Raman power.

## VI. Experiments

In this section we describe the experimental setup and the experimental results obtained for the silicon nitride waveguides with and without integrated nano-antennas, and compare with the theoretical models described in sections IV and V.

**A brief description of the experimental setup**
The details of the setup have been discussed elsewhere [**24, 27**], but for convenience of the readers, the setup is reproduced in Fig. 6, and briefly described below. A tunable Ti:sapphire laser is set to a pump wavelength of $\lambda_P = 785$ nm (red) after which the polarized beam passes through a half-wave plate ($\lambda_0/2$) in order to rotate the polarization to a TE-polarized beam. The pump beam then passes through a laser line filter (LLF) for side-band suppression before it is coupled into the chip by an aspheric lens (ASPH). The output beam is collected with an objective (OBJ) and passes through a polarizer P (set to TE) before it is filtered by a dichroic mirror which reflects the pump beam and transmits all Stokes wavelengths (green). The Stokes light is collected into a fiber using a parabolic mirror collimator (PMC) after which the fiber is split by a fiber splitter (FS) of which 1% goes to a power meter (PM) and 99% to a commercial spectrometer from ANDOR (Shamrock 303i spectrometer and iDus 416 cooled CCD camera). The 1% fiber tap is used during alignment and to measure the transmitted power.

**Experiments with waveguides without nanoantennas**
The analyte, which is pure isopropyl alcohol (IPA) here, is drop-casted on the chip containing the waveguides and covers the relevant waveguide region. The Raman signal of the analyte is collected via the same waveguide and measured with a commercial spectrometer as described in the previous section. For strip waveguides, a width $w$ of 700 nm is chosen while for slot waveguides, a slot width $s = 150$ nm, and total waveguide width $w = 700$ is chosen. Both of the waveguides have a height $h = 220$ nm. The waveguides are 1cm in length and are coiled as spirals (typical footprint: 800 μm x 500 μm).

Fig. 7(a) shows the background observed from the waveguide before application of IPA and the Raman spectrum of IPA observed on it after application of IPA. The background was removed by subtracting the truncated polynomial fit to the background [**45**]. Fig. 7(b) shows typical Raman spectra of IPA obtained using such a procedure for a strip waveguide ($w = 700$ nm) and a slotted waveguide ($w = 700$ nm and $s = 150$ nm). We can clearly distinguish all the major peaks of IPA obtained using both waveguide types. As an alternate analyte we also apply a 1M glucose solution on top of the slot waveguide ($w = 700$ nm, $s = 150$ nm). In Fig. 7 (c) the spectra obtained from the glucose solution before and after background correction are shown. We observe strong peaks that match well with the known Raman spectrum of α-D glucose solution.

One key highlight of Fig 7(b) is that about an eight-fold increase in the signal is observed for slotted waveguides compared to the strip waveguides, in accordance with the theory. To verify the theory further, in Fig. 8, the experimentally determined values of $\eta_0$ for various waveguide geometries and polarizations are compared with the theoretical values obtained using Eq. (6) and the signal corresponding to the 819 cm$^{-1}$ vibration of IPA. The theoretical curves of $\eta_0$ as a function of waveguide width $w$ are found to be within the error margin of the experimental data for both polarizations and the waveguide types studied. These experimental results validate the theoretical model discussed in Sec IV.

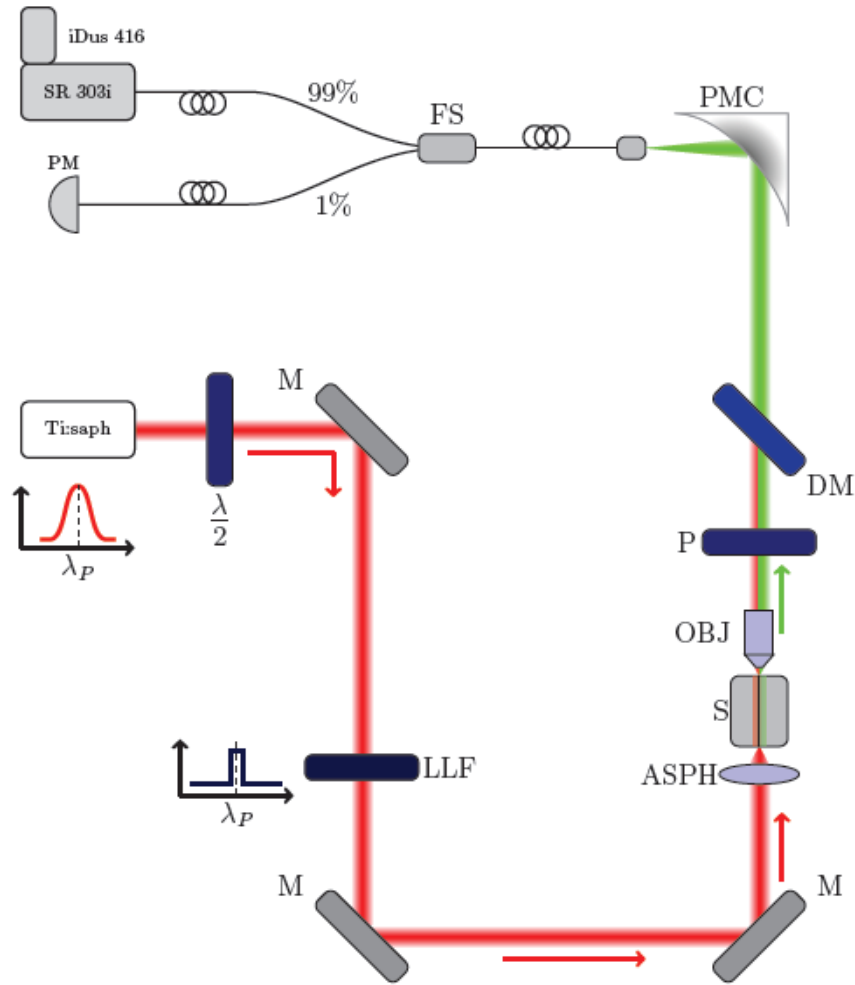

Fig. 6: Ti:saph: tunable Ti:saphire laser emitting the pump beam at $\lambda_P = 785$ nm, PM: power meter, SR 303i and iDus 416: spectrometer and cooled CCD detector from ANDOR, BS: beamsplitter, $\lambda_0/2$: half-wave plate, LLF: laser line filter for 785 nm, P: polarizer, M: fixed mirror, OBJ: objective (50X, NA=0.9), ASPH: aspheric lens (NA=0.5), S: sample stage, DM: dichroic mirror, PMC: parabolic mirror collimator (EFL=15 mm, NA=0.2), FS: fiber splitter (adapted from [27]).

The reported experiments and simulations are performed for IPA as the analyte in the uppercladding. However, the experimental and theoretical assessments have also been performed using other liquid and solid analytes, such as, acetone, dimethyl sulfoxide, and spin coated photoresists. The confinement properties of the waveguide mode changes if the refractive index of the uppercladding varies; therefore, according to Eq. (6) the numerical values of the conversion efficiency will slightly differ. Nonetheless, the results discussed in this article are pertinent for analytes of general interest such as a glucose solution.

Next we discuss experimental results relating to the integration of plasmonic nano antennas on top of the waveguides.

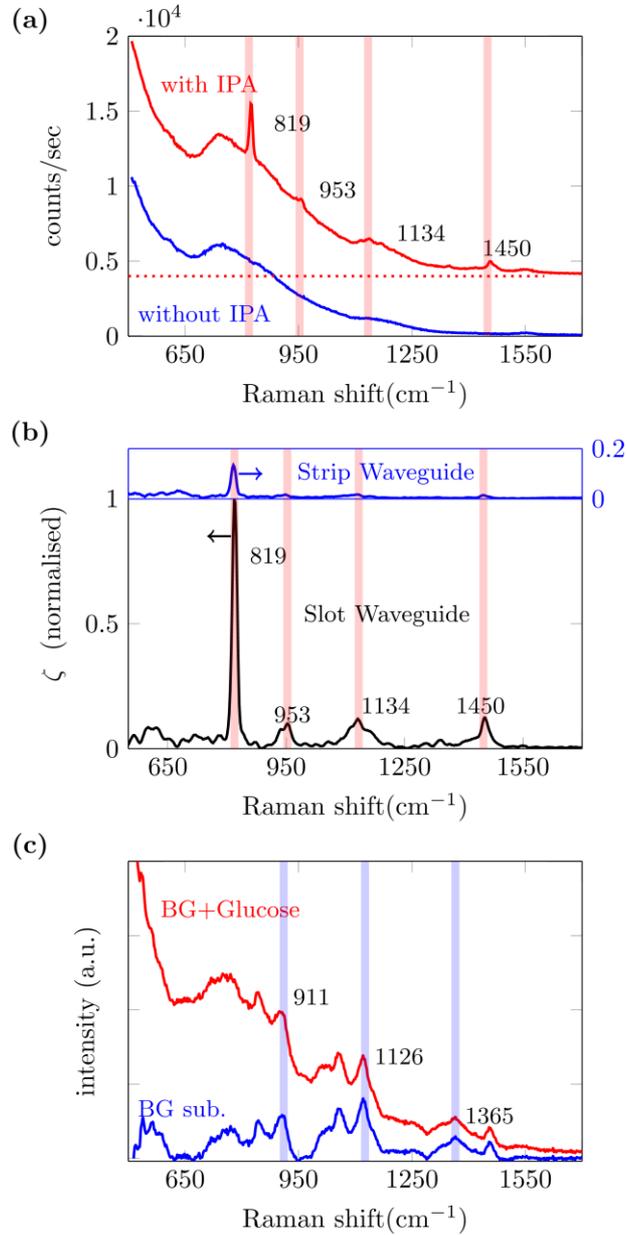

Fig. 7. (**a**) Raman spectra measured from a 1.6 cm waveguide ($w$ = 700 nm) without IPA (blue) and with IPA (red) on top. The spectrum with IPA is shifted vertically with zeros at the dashed red line (adapted from [**24**]). (**b**) Evanescently measured Raman spectra of IPA after background subtraction. The spectra are normalized to the 819 cm$^{-1}$ peak of obtained from the slot waveguide ($w$ =700 nm, $s$ = 150 nm). The blue spectrum (with blue axes) is obtained by using a 700 nm wide $Si_3N_4$ strip (adapted from [**25**]). (**c**) Evanescently measured Raman spectra of 1M glucose solution before and after background subtraction obtained from the slot waveguide ($w$ =700 nm, $s$ = 150 nm).

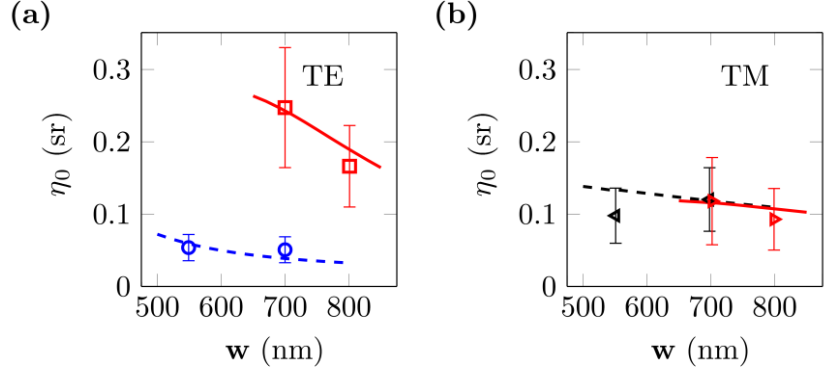

Fig. 8. The theoretical and experimental values of $\eta_0$ obtained for (**a**) TE modes and (**b**) TM modes of $Si_3N_4$ waveguides. The markers with error bars represent the estimated experimental errors [24]. The lines represent theoretical curves. The red solid lines are the theoretical curve for slot waveguides with s = 150 nm. The blue and black dashed lines are the theoretical curves for TE and TM polarizations respectively for strip waveguides. Circle: TE polarization, strip waveguides. Square: TE polarization, slot waveguides. Left handed triangles: TM polarization, strip waveguides. Right handed triangles: TM polarization, slot waveguides (adapted from [**25**]).

**Experiments with hybrid nanophotonic-plasmonic waveguides**

The nanoantennas were patterned on top of $Si_3N_4$ waveguides as described in [**27**]. The antennas were coated with 4-nitrothiophenol (NTP), which is assumed to form a self-assembled monolayer on the gold through a Au-S bond and which acts as a probe molecule to detect SERS events. Since NTP does not bind to $Si_3N_4$ this guarantees that any Raman signal has a SERS nature. The fundamental TE mode of a $Si_3N_4$ rib waveguide excites the SERS signal, which is subsequently collected back by the same waveguide mode.

Figure 9 depicts the Raman spectra, before and after coating with NTP, of waveguides functionalized with varying $N$=10, 20, 30, 40 (for a fixed bowtie geometry with $\alpha$ = 60°, $\Delta$ =48 ± 13 nm L= 106± 8). The spectrum of a reference waveguide ($N=0$) is shown as well. Each Raman spectrum is averaged 10 times. The spectral regions where an NTP Stokes peak is expected (1080, 1110, 1340 and 1575 $cm^{-1}$) [**27**] are highlighted by the cyan shaded areas. The peaks at 1250 and 1518 $cm^{-1}$ (marked by the black dashed lines) are already visible before coating, so they are not due to the NTP but are attributed to interference effects of the Au array [**26, 27**]**.** After coating additional peaks, which perfectly coincide with the expected NTP Stokes peaks, appear. While the non-functionalized reference waveguide does not generate any NTP SERS signal, it does generate a relatively large Raman background originating from the $Si_3N_4$ waveguide core (with a length of 1 cm). The shot noise associated to this background will limit the detection of the smallest NTP Raman features. This is evidenced by the fact that the smallest 1110 $cm^{-1}$ peak only appears for the $N$ = 40 waveguide, which has a reduced background as compared to the $N$ = 10 case due to the attenuation caused by the larger number of antennas. However, the absolute signal strength of the $N$ = 40 peaks is lower than the $N$ = 10 peaks as shown in Fig. 10 where the signal strength and Signal-to-Noise Ratio (SNR) of the 1340 $cm^{-1}$ peak is plotted for different $N$. The solid lines represent a fit of the experimental data to our earlier described on-chip SERS model. The shaded areas represent a distribution of possible signal counts and SNR values that arise due to fabrication induced variations among all antennas in the array. Additional details can be found in [27].

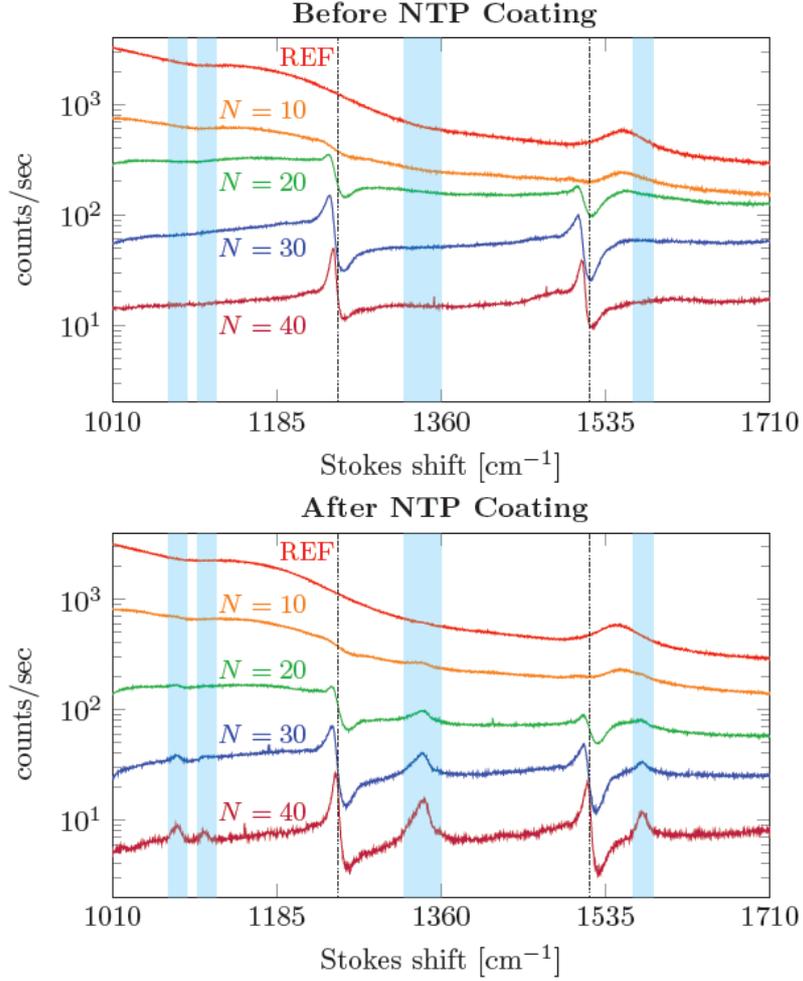

Fig. 9: Raman spectra, before and after NTP coating, of a reference waveguide (*N*=0) and waveguides functionalized with *N=10, 20, 30, 40* antennas (adapted from [27]).

Furthermore one can see that the signal is maximized for *N=N$_{opt}$*, while the SNR is maximal for $N = N_{opt}^{SNR} > N_{opt}$. It is expected that background mitigation through proper chip designs, and not through the antennas itself, can significantly enhance the SNR because one maintains the possibility of optimizing the signal while reducing the background simultaneously. From Fig. 10 it moreover follows that a minimum number of antennas *N$_{min}$* is required to achieve an SNR>1. Proper background mitigation is also expected to push towards the *N$_{min}$* = 1 limit such that signals from a single antenna can be detected.

For the data shown in Fig. 9, we obtained a fitted value of $\eta_A = 2.6 \pm 0.77 \cdot 10^{-15}$ which was in excellent correspondence with the theoretically predicated value $\eta_A = 2.2 \cdot 10^{-15}$ as detailed elsewhere [27]. This clearly establishes the validity of our model and its ability to provide quantitative predictions of the absolute Raman power coupled into a single mode waveguide. For this particular case, a single antenna, coated with an NTP monolayer, will hence produce 2.6 ± 0.77 fW for 1 W of guided power.

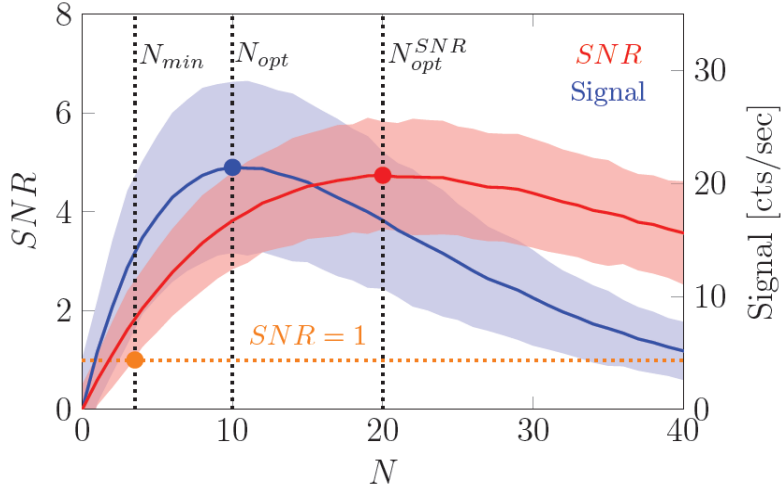

Fig. 10: Signal and SNR of the 1340 cm$^{-1}$ peak. The solid lines and shaded areas represent a fit to our on-chip SERS model (adapted from [**27**]).

### VII. Comparative performance analysis of waveguides with and without nanoantennas

For the case of evanescent waveguide sensing without nanoantennas, the signal enhancement is mostly due to the large interaction volume distributed over the length of the waveguide. Hence, these sensors are suitable for sensing relatively larger analyte volumes. The enhancement of these sensors cannot be increased arbitrarily and a very high enhancement, needed for e.g. single molecule detection, is not possible using standard designs.

However, for the waveguides with integrated nanoantennas, the enhancement is the result of a plasmonic resonance effect at discrete 'hot-spots', i.e. regions of extreme localization of the electric field. Hence, waveguides integrated with nanoantennas are suitable for studying very small number of localized molecules. We have theoretically shown that for integrated nanoantennas, an enhancement as high as $10^{10}$ is possible [**46**]. Hence, single molecule detection in an integrated framework is possible with such hybrid nanophotonic-plasmonic waveguides.

Depending on the dimensions and the geometry of the antennas and the waveguide, we estimated that each integrated antenna discussed in this paper has an equivalent $Si_3N_4$ rib waveguide length of 250-2500 nm. This equivalent length is the physical length of a bare $Si_3N_4$ waveguide (without any gold) that is required to generate the same signal as one integrated nanoplasmonic antenna. As a rough rule of thumb, 1 $\mu m$ of a $Si_3N_4$ rib waveguide produces the same Raman signal as 1 plasmonic nanoantenna integrated on top of the waveguide. However, for the nanoantennas, the loss associated with gold absorption, elastic scattering, surface roughness etc. is about $e_p \sim e_s \sim 0.5$ dB per antenna [**27**]. The loss per antenna is quite high compared to the $10^{-4}$ dB/$\mu m$ for waveguides without the antenna. For bulk sensing, the integration of nanoplasmonic antennas is hence completely superfluous because the signal generated by bare waveguides will be much larger due to the large waveguide length and small waveguide loss. On-chip SERS will however outperform the bare waveguide sensing case when particular analytes are adsorbed on the metal surface. In that case one can make use of the extreme local Raman enhancement of the plasmonic hotspots, which can be orders of magnitude higher than the field enhancement of dielectric rib or slot waveguides.

As discussed in section IV, depending on the waveguide geometry, more than 3 to 5 orders of magnitude higher signal is expected from a waveguide (bare) compared to diffraction-limited systems. Indeed, for a realistic silicon nitride waveguide ($h = 220$ nm, $s = 150$ nm, $w = 700$ nm, $l = 1$ cm) about three orders of magnitude is obtained experimentally [**29, 47**]. As a consequence of this enhancement, a higher signal-to-noise ratio (SNR) is obtained compared to the detector dark-noise-limited system, despite the background noise from the waveguide core [**47**]. Still higher performance is expected for a very thin layer of molecules functionalized on top of the chips. It results from the fact that the signal from microscopes is limited by the thickness $t$ of the layer, so that $z_R$ in Eq. (1) is replaced by $t$. Whereas for waveguide systems, the signal essentially remains the same, as the strongest contribution for $\eta_0$ in Eq. (7) comes from the molecules near the waveguides in either cases. More than four orders of magnitude signal enhancement is expected for a very thin layer using a centimeter of waveguide length.

Our on-chip SERS platform improves on existing hybrid nanophotonic circuits where waveguides are used to probe SERS signals from external metallic nanoparticles, in the sense that it allows a quantitative prediction of the Raman enhancement and coupling with the underlying waveguide. The complete integration moreover allows, by fabrication, a robust and optimized alignment and can potentially improve the coupling efficiency between the excitation beam and the nanoplasmonic antennas as compared to free-space excitation due to the near-field interaction of the guided mode and the plasmonic mode [**48**]. Currently we don't gain in terms of maximum Raman enhancement factor (about $10^4$) as compared to existing technologies. It is however expected that by decreasing the gap size of the bowtie antennas, the maximum Raman enhancement can be boosted by another two to three orders of magnitude.

### VIII. Conclusion

We reviewed an integrated waveguide approach for on-chip spontaneous and surface-plasmon enhanced Raman spectroscopy using the evanescent field of the dielectric waveguide for excitation and collection. The effects of the waveguide parameters and polarization of the modes on the overall evanescent excitation and collection efficiency of spontaneous Raman scattering were discussed and validated experimentally. Thanks to the use of low-loss, high index contrast waveguides, more than two orders of magnitude improvement in the overall excitation and collection efficiency is estimated for a centimeter of waveguide as compared to the free space excitation and collection. Moreover we developed an analytical model for on-chip SERS to identify the key design parameters and figure of merit for a hybrid nanophotonic-plasmonic platform. The experimentally obtained SERS signals were in excellent correspondence with our theoretical analysis. Depending on the number of analyte molecules to be probed, the integrated waveguide approach provides alternate routes for Raman spectroscopy. We conclude that 'bare' waveguides perform better when a large numbers of molecules have to be probed. Integrated nanoantennas are particularly useful when analyte adsorb to the metal surface in which case the Raman signal can be boosted by several orders of magnitude due to the plasmonic enhancement. Moreover on-chip SERS could be recommended for probing extremely weak Raman scatterers or ultralow concentrations provided the analyte molecules find their way to the plasmonic hot spots. This enables a comprehensive design freedom and quantitative control on the Raman signal acquisition depending on various applications. In combination with other on-chip devices, such as sources, filters and arrayed waveguide gratings, the presented on-chip Raman platform allows multiplexed detection of extremely weak Raman signals on a highly dense integrated platform.

**Acknowledgements**

The authors acknowledge the ERC advanced grant InSpectra for partial funding, and imec, Leuven for fabrication of the waveguides.